\documentclass[a4paper]{article}

\usepackage{INTERSPEECH2016}

\usepackage{graphicx}
\usepackage{amssymb,amsmath,bm}
\usepackage{textcomp}
\usepackage{array}
\usepackage{stfloats}

\ninept

\title{Max-Margin Metric Learning for Speaker Recognition}

\makeatletter
\def\name#1{\gdef\@name{#1\\}}
\makeatother \name{{\em Lantian Li, Dong Wang, Chao Xing, Thomas Fang Zheng$^*$\thanks{{This work was supported by the National Natural Science Foundation of China under Grant No. 61371136 and No. 61271389, it was also supported by the National Basic Research Program  (973 Program) of China under Grant No. 2013CB329302. D.W. and T.F.Z. are with Division of Technical Innovation and Development of Tsinghua National Laboratory for Information Science and Technology and Research Institute of Information Technology (RIIT) of Tsinghua University. This paper is also supported by Pachira.}}}}

\address{Center for Speech and Language Technologies, Tsinghua University \\
    {\small \tt \{lilt,xingchao\}@cslt.riit.tsinghua.edu.cn; wangdong99@mails.tsinghua.edu.cn} \\
    {\small \tt $^*$Corresponding Author:fzheng@tsinghua.edu.cn}}

\begin{document}

\maketitle
\begin{abstract}

Probabilistic linear discriminant analysis (PLDA) is a popular normalization approach for the i-vector model,
and has delivered state-of-the-art performance in speaker recognition. A potential problem of the PLDA model,
however, is that it essentially assumes Gaussian distributions over speaker vectors, which is not always
true in practice. Additionally,
the objective function is not directly related to the goal of the task, e.g., discriminating true speakers
and imposters.

In this paper, we propose a max-margin metric learning approach to solve the problems. It learns a linear transform
with a criterion that the margin between target and imposter trials are maximized.
Experiments conducted on the SRE08 core test show that compared to PLDA, the new approach can
obtain comparable or even better performance, though the scoring is simply a cosine computation.

\end{abstract}

\section{Introduction}

The i-vector model represents the state-of-the-art architecture of modern speaker recognition~\cite{ES3,ES4}. By this model,
a speech segment is represented as a low-dimensional continuous vector (i-vector), so that speaker recognition
(and other tasks) can be performed based on the vector representations.

A particular property of the i-vector model is that both the speaker and session variances are embedded in a
single low-dimensional subspace.
This is an obvious advantage since more speaker-related information is retained compared
to other factorization models, e.g., JFA~\cite{ES3}; however, since the speaker-related information is
buried under others, raw i-vectors are not sufficiently discriminative with respect to speakers.
In order to improve the discriminative capability of i-vectors for speaker recognition, various normalization or
discrimination
models have been proposed, including within-class covariance normalization (WCCN)~\cite{hatch2006within},
nuisance attribute projection (NAP)~\cite{solomonoff04}, linear discriminant analysis (LDA)~\cite{ES1}, and
its Bayesian counterpart, probabilistic linear discriminant analysis (PLDA)~\cite{ES6}.

Among these models, PLDA plus length normalization is reported to be the most
effective.
The success of this model is largely attributed to two factors:
one is its training objective function that reduces the intra-speaker variation while
enlarges inter-speaker variation, and the other is the Gaussian prior that is assumed over speaker vectors,
which improves robustness when inferring i-vectors for speakers with limited training data.

These two factors, however, are also the two main shortcomings of the PLDA model.  As for the objective function,
although it encourages discrimination among speakers, the discrimination is based on Euclidian distance, which is
inconsistent with the normally used cosine distance that has been demonstrated to be more effective in speaker recognition.\footnote{This inconsistency is more serious for the LDA model
for which cosine distance is used in evaluation. For PLDA, the training and evaluation are both with
Euclidian distance so there is not much inconsistency. However, since cosine distance is potentially
more suitable in speaker recognition,
a model purely based on cosine distance is preferred.} Additionally, our task in speaker recognition
is to discriminate true speakers and imposters,
which is a binary decision rather than the multi-class discrimination in PLDA training. As for the Gaussian
assumption,
although it greatly simplifies the model, the assumption is rather strong and is not very practical in
some scenarios, leading to less representative models.



Some researchers have noticed these problems. For example, to go beyond the Gaussian assumption, Kenny~\cite{kenny2010bayesian} proposed
a heavy-tailed PLDA
which assumes a non-Gaussian prior over the speaker mean vector. Garcia-Romero et al.~\cite{garcia2011analysis} found that length normalization can
compensate for the non-Gaussian effect and boost performance of Gaussian PLDA to the level of the heavy-tailed PLDA.
Burget, Cumani and colleagues~\cite{burget2011discriminatively,cumani2013pairwise} proposed
a pair-wised discriminative model that discriminates true speakers and imposters.
In their approach, the model accepts a pair of i-vectors and predicts the probability that they belong to the same speaker.
The input features of the model are derived from the i-vector pairs according to a form derived from the PLDA score function (further generalized to any symmetric
score functions in~\cite{cumani2013pairwise}), and the model is trained on i-vector pairs that have been labelled as identical
or different speakers. A particular shortcoming of this approach is that the feature expansion is highly complex. To solve this problem,
a partial discriminative training approach was proposed in~\cite{6936581}, which optimizes the discriminative model on a
subspace and does not require any feature expansion. In~\cite{wang2014discriminative}, we proposed a discriminative approach
based on deep neural networks (DNN), which holds the same idea as the pair-wised training,
whereas the features are defined manually.

Although promising, the discriminative approaches mentioned above seem rather complex. We hope a model as simple as LDA and the scoring is as simple as a cosine computation. This paper presents
a max-margin metric learning (MMML) approach, which is a simple linear projection trained with the objective of discriminating
true speakers and imposters directly. Once the projection has been learned, simple cosine distance is sufficient to conduct the scoring.
This approach belongs to the simplest metric learning methods which have been studied for decades in machine
learning~\cite{ES12,ES14}, though it has not been extensively studied in speaker recognition.

The rest of this paper is organized as follows. Section~\ref{sec:rel} discusses some related work, Section~\ref{sec:cdml} presents
the max-margin learning method. The experiments are presented in Section~\ref{sec:exp}, and Section~\ref{sec:conl}
concludes the paper.

\section{Related work}
\label{sec:rel}

Some of the related studies, particularly the pair-wised discriminative model, have been discussed in the previous section. This
section presents some studies on metric learning for speaker recognition, which are related to our study more directly.
A representative work proposed in~\cite{ES11}
employs neighborhood component analysis (NCA) to learn a projection matrix that minimizes the average leave-one-out k-nearest neighbor classification error.  Our model differs
from the NCA approach in that we use max-margin as the training objective and cosine distance as the similarity measure, which is more suitable for speaker recognition.

The cosine similarity large margin nearest neighborhood (CSLMNN) model proposed in~\cite{ES5} is more relevant to our proposal. The authors
formulated the training task as a semidefinite program (SDP)~\cite{weinberger2005distance} which moves i-vectors of the same speaker closer by maximizing the cosine distance among them, while separating i-vectors of different speakers by a large margin.
Our approach uses a similar objective function, but employs a simpler solver based on stochastic gradient descent (SGD), which supports mini-batch learning and accommodates large-scale optimization.


\section{Max-margin metric learning}
\label{sec:cdml}

This section presents the max-margin metric learning for speaker recognition. Metric learning
has been studied for decades. The simplest form is to learn a linear projection $M$ so that the distance among the projected
data is more suitable for the task in hand~\cite{ES12}. For speaker recognition, the most popular used distance metric is
the cosine distance and the goal is to discriminate true speakers and imposters. We therefore optimize $M$ to make the projected
i-vectors more discriminative for genuine and counterfeit speakers measured by cosine distance.

Formally, the cosine distance between two i-vectors $w_1$ and $w_2$ is given as follows:

\vspace{-1mm}
\begin{equation}
   d(w_1,w_2) = \frac{<{w}_1,{w}_2>}{\sqrt{||{w}_1|| ||{w}_2||}}.
\label{eq1}
\end{equation}

\noindent where $<\cdot,\cdot>$ denotes the inner product, and $||\cdot||$ is the $l$-2 norm. Further define a contrastive triple $(w,w^+,w^-)$
where the two i-vectors $w$ and $w^+$ are from the same speaker, and $w$ and $w^-$ are from different speakers. Letting $S$ denote all the contrastive triples in a development set, we can define the max-margin objective function that encourages i-vectors
of the same speaker moving close while driving i-vectors from different speakers far away. This is formulated as follows:

\vspace{-1mm}
\begin{equation}
    \mathcal{L}(M) = \sum_{(w,w^+,w^-) \in S} \max \{ 0, \delta - d^+(M) + d^-(M)\}
\label{eq2}
\end{equation}

\noindent where $\delta$ is a hyperparameter that determines the margin, and

\[
d^+(M)= d(M w, M w^+)
\]
\[
d^-(M)= d(M w, M w^-).
\]

\noindent Note that minimizing this function results in  maximizing the margin between i-vectors of the same speaker and different speakers.

Note that optimizing $\mathcal{L}(M)$ directly is often infeasible, because the size of $S$ is exponentially large. We choose the SGD
algorithm to solve the problem, where the training is conducted in a mini-batch style. In a mini-batch $t$, a number of contrastive
triples are sampled from $S$, and these triples are used to calculate the gradient $\frac{\partial{\mathcal{L}}}{\partial{M}}$.
The projection $M$ is then updated with this gradient as follows:

\vspace{-1mm}
\begin{equation}
  M^{t} = M^{t-1} + \epsilon \frac{\partial{\mathcal{L}}}{\partial{M}}
\label{eq3}
\end{equation}
\noindent where $M^t$ is the projection matrix at mini-batch $t$, and $\epsilon$ is a learning rate. This learning iterates until
convergence is obtained. In this study, the Theano package~\cite{bergstra2010theano} was used to implement the SGD training.

Once the matrix $M$ has been learned from the development data, an i-vector $w$ can be projected to its image $M w$
in the projection space, where true speakers and imposters are more easily to be discriminated, according to the training objective.
Note that the max-margin metric learning is based on cosine distance, which means that the simple cosine distance
is the theoretically correct choice when scoring trials in the projection space. This is a big advantage compared to PLDA that requires complex matrix computation.

\begin{figure*}[pt]
   \centering
   \vspace{-1mm}
   \includegraphics[width=12cm]{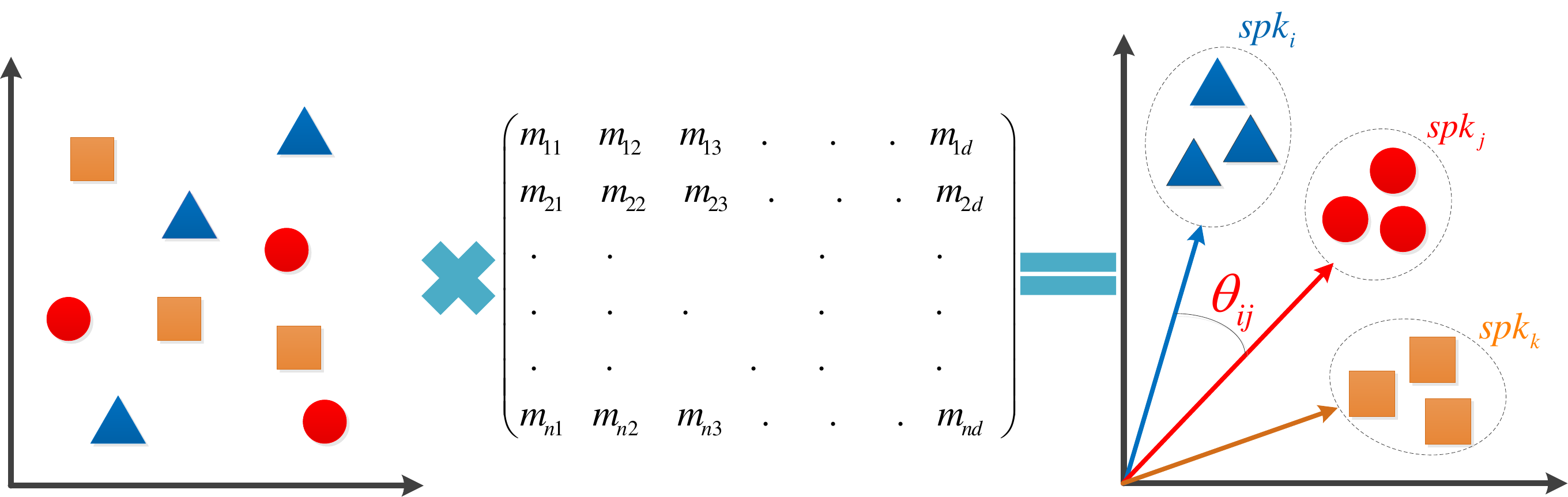}
   \caption{{\it Illustration of the improved discrimination with the max-margin metric learning. Each speaker is represented by a
   shape and a particular color. After applying the projection that is learned from data, speakers that congest together in the original  i-vector space become separated.}}
   \label{fig:metric}
\end{figure*}

Figure \ref{fig:metric} illustrates the concept of the max-margin metric learning for speaker recognition. The i-vectors from the same speaker are
labeled as the same color and shape. In the input space, i-vectors of all the speakers are congested together. After applying
the learned projection, i-vectors of the same speaker are moved closer, while those of different speakers are moved apart. Note that $\theta_{ij}$ is the margin between the speaker pair $spk_i$ and $spk_j$.

\section{Experiments}
\label{sec:exp}

We evaluate the proposed method on the SRE08 core test. This section first presents the data used and the experimental setup,
and then reports the results in terms of equal error rate (EER) and DET curves.

\subsection{Database}

The Fisher database was used to train the i-vector system. We selected $7,196$ speakers to train the i-vector model, the LDA model and the PLDA model.
The same data were also used to conduct the metric learning. The NIST SRE 2008 evaluation database~\cite{ES9} was used as the test set. We selected $1,997$ female utterances from the core evaluation data set (short2-short3) and based on
which constructed $59,343$ trials, including $12,159$ target trials and $47,184$ imposter trials.

\subsection{Experimental setup}

The basic acoustic features involved 19-dimensional Mel frequency cepstral coefficients (MFCCs) and the log energy.
This basic features were augmented by their first and second order derivatives, resulting in 60-dimensional
feature vectors.
The UBM involved $2,048$ Gaussian components and was trained with about $8,000$ female utterances selected from the Fisher database randomly.
The dimensionality of the i-vectors was $400$. The LDA model was trained with utterances of $7,196$ female speakers, again randomly selected from the Fisher database.
The dimensionality of the LDA projection space was set to $150$.
For the metric learning, utterances in the Fisher database were sampled randomly to build the contrastive triples and were used to train the projection matrix.

\subsection{Basic results}
\label{sec:sv-task}

We first present the basic results obtained with various discriminative models: raw i-vectors with cosine scoring (Cosine), LDA, PLDA,
max-margin metric learning (MMML). The test is based on the NIST SRE 2008 core task, which is divided into $8$ test conditions according to the channel,
language and accent~\cite{ES9}. The EER results are reported in Table~\ref{tab:exps}.

It can be observed that the proposed MMML significantly improves the discriminative capability of raw i-vectors, and it
outperforms both LDA and PLDA in condition $1$-$4$ (which takes the major proportion of the test data). In condition $5$-$8$,
the PLDA wins the competition. We attribute this discrepancy to the data imbalance in the development set:
condition $5$-$8$ involves complex patterns (e.g., multilingual speakers, different accents) that were not involved in the
Fisher database, which is used to train the MMML model. This leads to performance degradation on these conditions
with the MMML approach. For LDA and PLDA, the Gaussian assumption improves
generalizablility on unseen conditions, thus resulting in superior performance than MMML, a purely discriminative approach.
Nevertheless, since condition $1$-$4$ takes a large proportion of the data, the MMML approach gets the best overall performance. As a summary, it seems that the MMML model requires more data to cover complex acoustic conditions,
whereas the PLDA model is more robust to data variance, due to its prior Gaussian assumption.

\vspace{-2mm}
\begin{table}[htp]
        \centering
        \caption{\it EER results on NIST SRE 2008 core test. The best results are shown in bold face for each condition.}
        \vspace{2mm}
          \begin{tabular}{|l|c|c|c|c|}
           \hline
                             &\multicolumn{4}{c|}{EER\%}\\
           \hline
                 Condition   &   Cosine   &   LDA    &   PLDA        &   MMML          \\
           \hline
                   C1        &   26.93    &  20.80   &   17.64       &  {\bf 14.13}     \\

                   C2        &    4.48    &  1.79    &    2.09    &  {\bf 1.19 }      \\

                   C3        &   26.71    &  20.97   &   17.56       &  {\bf 14.69 }   \\

                   C4        &   19.22    &  13.21   &   13.96       &  {\bf 11.11 }    \\

                   C5        &   19.71    &  14.78   &   {\bf 11.30} &  12.02         \\

                   C6        &   11.31    &  9.92   &   {\bf 8.04}   &  10.64          \\

                   C7        &   7.22     &  5.58    &   {\bf 4.18}  &  6.34          \\

                   C8        &   7.37     &  6.32    &   {\bf 4.74}  &  6.05           \\
            \hline
                Overall      &   23.88    &  20.24   &   17.95       &  {\bf 15.31}    \\
            \hline
          \end{tabular}
          \label{tab:exps}

\end{table}

\begin{figure}[!htb]
   \centering
   \includegraphics[width=8cm]{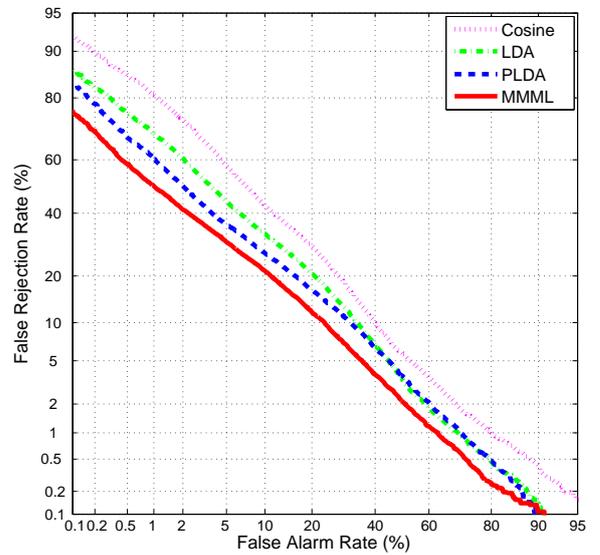}
   \caption{{\it The DET curves on the NIST SRE 2008 overall test condition.}}
   \label{fig:det}
\end{figure}

The DET curves on the overall test condition with the four models are presented in Figure ~\ref{fig:det}. It is clearly observed that the MMML approach outperforms the other three methods.

\vspace{-2mm}

\begin{table}[!htp]
        \centering
        \caption{\it EER results with LDA/MMML tandem composition.}
        \vspace{2mm}
          \begin{tabular}{|l|c|c|}
           \hline
                             &\multicolumn{2}{c|}{EER\%}\\
           \hline
                 Condition   &  MMML + LDA   & LDA + MMML    \\
           \hline
                   C1        &  19.84      &   14.16   \\

                   C2        &  2.09       &   1.49   \\

                   C3        &  19.96      &   14.69   \\

                   C4        &  12.16      &   10.66   \\

                   C5        &  14.90      &   11.66   \\

                   C6        &  9.98       &   10.37   \\

                   C7        &  5.70       &   6.34   \\

                   C8        &  6.32       &   6.05   \\
            \hline
                Overall      &  19.68      &   15.25   \\
            \hline
          \end{tabular}
          \label{tab:tandem}
\end{table}

\subsection{Tandem composition}

We note that both LDA and MMML learn a linear projection, though they are based on different
learning criteria: LDA uses Fisher discriminant while MMML uses max-margin.
The results in Table~\ref{tab:exps} show that the max-margin criterion is clearly superior.
An interesting question is whether the two criteria can be composed in a tandem way.
The results are shown in Table~\ref{tab:tandem}, where the system `LDA+MMML' involves a $400 \times 150$
dimensional LDA projection followed by a $150 \times 150$ dimensional MMML projection, while the system `MMML+LDA' involves a $400\times150$ dimensional MMML and
a $150 \times 150$ dimensional LDA.  From these results, we find that the \emph{last} projection is
the most important: if it is an MMML, the performance is always good. The `MMML+LDA' system seems a bit
superior than the original LDA, which perhaps because the advantage of the max-margin training has been
consolidated in the process of dimension reduction, which benefits the subsequent LDA.

\subsection{Score fusion}

The LDA/PLDA model and MMML model are complementary: LDA/PLDA are generative models and
better generalizable to rare conditions where little training data are available, whereas
MMML is purely discriminative and is superior for matched conditions. Combining these two
types of models may offer additional gains. We experimented with a simple score
fusion approach that linearly interpolates the scores from LDA/PLDA and MMML.
It can be represented as $\alpha S_{MMML} + (1-\alpha) S_{LDA/PLDA}$, where $\alpha$ is the interpolation factor.
The results of the score fusion are presented in Table~\ref{tab:fusion}, where the
interpolation factor $\alpha$ is chosen to be $0.2$.
Compared to Table~\ref{tab:exps}, we observe that the fusion consistently leads to better performance
than the original LDA and PLDA systems. Interestingly, the performance
on condition $5$-$8$ is also improved,
although the MMML approach does not work well individually in these conditions. Note that the performance degradation
on condition $1$,$3$,$4$ compared to the original MMML system is simply because we use a global interpolation
factor. If the factor has been tuned for each condition separately, the fusion system
would obtain the best performance in all the conditions.

\begin{table}[htp]
        \centering
        \caption{\it EER results with score fusion, where the interpolation factor $\alpha$ is chosen to be $0.2$.}
          \vspace{2mm}
          \begin{tabular}{|l|c|c|}
          \hline
                             &\multicolumn{2}{c|}{EER\%} \\
           \hline
                 Condition   &   LDA + MMML    &   PLDA + MMML    \\
           \hline
           	       C1        &   18.11         &    15.88	\\

                   C2        &   1.49          &     1.79	\\

                   C3        &   18.48         &    15.97	 \\

                   C4        &   11.71         &    12.31    \\

                   C5        &   13.34         &    10.34     \\

                   C6        &   9.65          &    8.04      \\

                   C7        &   5.45          &    4.18	  \\

                   C8        &   5.53          &    4.47     \\
            \hline
                Overall      &   18.76         &   16.96     \\
            \hline
          \end{tabular}
          \label{tab:fusion}

\end{table}

\section{Conclusions}
\label{sec:conl}

In this paper, we proposed a max-margin metric learning approach for speaker recognition. This approach is a simple linear transform that is trained with the
criterion of max-margin between true speakers and imposters based on cosine distance. The scoring is as simple as LDA, but the performance is comparable or even better than
PLDA, especially with large training data on matched conditions. Future work will investigate metric learning with deep non-linear transforms, and
study better approaches to combine PLDA and MMML.

\newpage
\eightpt
\bibliographystyle{IEEEtran}

\bibliography{max-margin}

\end{document}